Furthermore, the ability to construct coherent structures appears to be of great potential value in a variety of practical problems. Over the past few years, techniques have been introduced for the control of low dimensional chaotic systems[20]. These techniques are limited by the dynamical dimension of the problem. That is, where the state of an entire system is described by only a few variables, one can hope to gain control of the system state. One cannot, however, either direct or constrain a cluster of particles in an extended chaotic flow, because each particle in the cluster is described by a distinct set of equations. Our recipe would, for the first time, allow us to hold such a cluster of particles together in an otherwise chaotic sea. This would have the dual effect of reducing the effective dynamical dimension of the particles in the cluster and enabling such applications as manipulating the trajectory of the entire cluster -- whose average position would now be described by a single set of equations -- *en masse*. In addition, it has not escaped our notice that structures are both prevalent and important in conventional problems in turbulence. For example, the methods described in this letter would seem to be appropriate tools for the creation and destruction of vortices trailing behind a bluff body immersed in a moving fluid.

The authors wish to thank D. Auerbach, L. Bresler, and G. Metcalfe for valuable comments. This work was supported by the Air Force Office of Scientific Research.

Let us now return to the questions raised at the beginning of this letter. As to the prevalence of coherent structures, we conclude that the folding of horseshoes, which is generic in chaotic problems, is in fact one cause of coherence in chaotic systems. These coherent regions explicitly require both stretching and folding and are consequently different from conventional elliptic islands in 2-D chaotic flows. We note as well that whereas the statistics of stretching are well characterized for common chaotic systems (distribution of Lyapunov exponents), no companion parameters for the quantification of folding exist as yet. We can, however, derive an approximate scaling relation for coherent structures in a (uncontrolled) discrete-time, spatially extended chaotic system as follows. To produce a coherent structure, a fixed point and a fold must be located within a distance, d, which is proportional to the width of the fold. This width is in turn determined by the thickness, $1/\Lambda$, (where $\Lambda$ is the Lyapunov number) of a horseshoe after one iterate of the map. Thus in D dimensions, the fixed point must be within the volume, $d^D \sim (1/\Lambda)^D$, encompassing a fold. The number of fixed points scales like $(1+\Lambda)$, while the number of folds scales like $\Lambda \cdot G(\Lambda)$, where $G(\Lambda)$ is a non-universal function which codifies how much folding occurs during each map iteration. So the number of folds within the distance d of a fixed point goes as $P(\Lambda) \sim \Lambda \cdot (1+\Lambda) \cdot G(\Lambda) \cdot \Lambda^{-D}$. Thus even in highly chaotic 2-D extended systems, the *a priori* expectation is that there should be at least one coherent structure (whose size, however, diminishes with $(1/\Lambda^2)$). In Fig. 5(a), for instance, the location of a small period two coherent structure is indicated by x's.

As to the persistence of coherent structures in chaotic dynamics, given that sufficiently compact folds near periodic points are a cause of coherence, it should be possible to destroy coherent structures either by displacing the periodic point or by disturbing the fold. Once this has been done, the underlying stretching of the chaotic system will destroy the structure. This has obvious implications for fluids problems such as transport or mixing, where coherent structures are to be avoided.

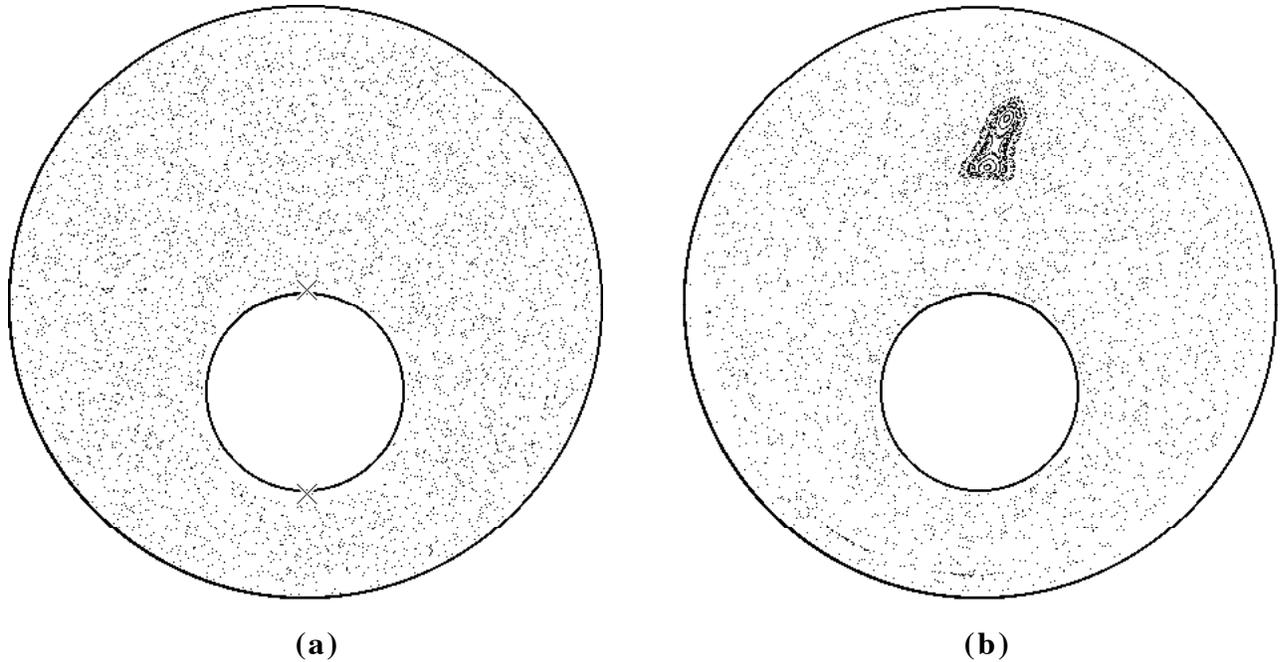

**Figure 5 - Marker particles for uncontrolled (a)
and control (b) rotation protocols**

Fig. 5 shows Poincaré sections corresponding to both the uncontrolled and the controlled protocols[15]. Marker particles starting inside the dark structure in Fig. 5(b) never leave the structure. Comparison of Fig's 4 and 5(b) reveals that the fold closer to P* in Fig. 4 leads to the central two elliptic regions, while the second fold coincides with the surrounding structure. By manipulating the fold locations, we are able to independently produce complete horseshoes, thereby destroying the structures resulting from either fold.

It thus appears that any system which exhibits horseshoes -- i.e. any chaotic system -- can in principle be manipulated to produce coherent structures. Computationally based flows -- e.g. tendril-whorl flows[16], egg-beater flows[17], cavity flows[18] -- are particularly simple and can be used to produce similar results in a straightforward manner. Other problems, such as the chaotic tangle produced behind a cylinder during periodic vortex shedding[19] necessitate further work, but are in principle amenable to the same analysis.

numerical experiments can be made identical[12]. By alternately driving the inner and outer cylinders at low speed, one can produce chaotic advection of fluid particles over most of the fluid volume. We consider the case in which the inner cylinder[13] is rotated by 1080°, then the outer by -240°, and the outer again by 1080°, where by convention we take clockwise rotation to be positive and counter-clockwise to be negative.

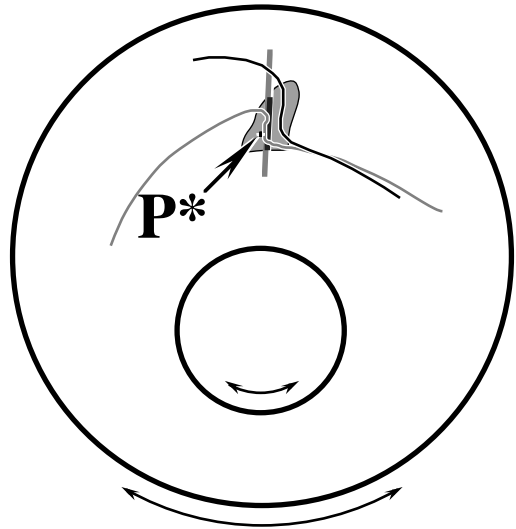

**Figure 4 - Fold created in chaotic fluid advection between two eccentric cylinders**

For this problem, it is possible to create new folds, shown in Fig. 4, near a pre-existing hyperbolic fixed point, P*. One can produce a coherent structure either by creating a new fold near a fixed point or by moving a pre-existing fold to a fixed point. We show the results only of creating new folds, although we have successfully tested both methods. . To produce the fold shown in Fig. 4, we change to the control protocol: outer: 32°, then inner: 1080°, then outer: -170°, and finally inner: 1080°. In the figure we display a line of initial points (grey) and their final locations after one (solid line) and two (dashed line) driving periods. Also shown is the resulting (cross-hatched) invariant region. In this system we are able to produce folds by manipulating global parameters; in other problems, local controls may be necessary[14].

This modified standard map repeatedly stretches and folds a segment of the initial line of points. The effect of the folds is to restrain points to an invariant neighborhood of the fixed point despite the fact that stretching is also present. This is shown in the inset of Fig. 3, where we display trajectories, starting from a small number of initial points near P*, under successive applications of the modified map. By contrast, without the added folds, the map [3] is chaotic in this region, and points starting in a neighborhood of P* quickly leave and travel globally in a large stochastic area.

We note that only points leading to the smaller of the two folds in Fig. 3 (closer to the original grey line of points) are constrained to this coherent structure. Points leading to the larger fold are mapped to distant regions, and are not coherent. This is an example of a failure to place the fold sufficiently close to the periodic point. The different behaviors of the two folds is notable because only the larger fold constitutes a well-formed, as opposed to incomplete[10], horseshoe. The incomplete horseshoe, by contrast, contains the fold and consequently produces the coherent structure.

We further note that, due to our particular choice of F(X), the fixed point, P*, in the center of Fig. 3 is hyperbolic. Eigenvalue analysis reveals that P* changes from elliptic to hyperbolic when $(dF(X)/dX)|_{P*}$ exceeds 4. Thus by locally increasing the slope of F(X) near P*, we can change the dynamics in the vicinity of the central fixed point from elliptic to hyperbolic -- all within a coherent structure which is itself embedded in a stochastic region. It is apparent that the coherence of the structure does not depend on the character of the fixed point, and no sort of conventional local analysis would be of any use at all in determining the ultimate fate of points within R.

The same idea can be applied to experimentally realizable systems. As an example, consider the problem of a fluid constrained between two asymmetrically placed cylinders (Fig. 4). This is one of the best studied examples of chaotic advection and has the virtue that its velocity field can be expressed analytically[11], so the results of physical and

*page 6*

where X and Y are defined on the periodic domain, $[0, 2\pi)$, and where K is a positive constant. This map has an unstable fixed point, $P^*$, at $X = \pi$, $Y = 0$, which is embedded in a stochastic region.

Let us introduce a fold in a region, R, around $P^*$ by replacing $K \cdot \sin(X_n)$ in Eq. [3] with a function, $F(X_n)$, when $(X_n, Y_n) \in R$. To produce a fold, we choose $F(X)$ so that it contains at least one extremum in R, and for continuity we require that $F(X) = K \cdot \sin(X)$ at the bounds of R. As an example, we choose $F(X)$ to be a modified cubic polynomial[9] with one minimum and one maximum in R: $2.81 < X_n < 3.41$, $-0.3 < Y_n < 0.3$. In Fig. 3, we show a line of initial points (in grey), and the first and second iterates of this line (as solid and dashed lines respectively) using the modified standard map with $K = 2.5$. Also shown as a cross-hatched area is the neighborhood of $P^*$ that is invariant under this map.

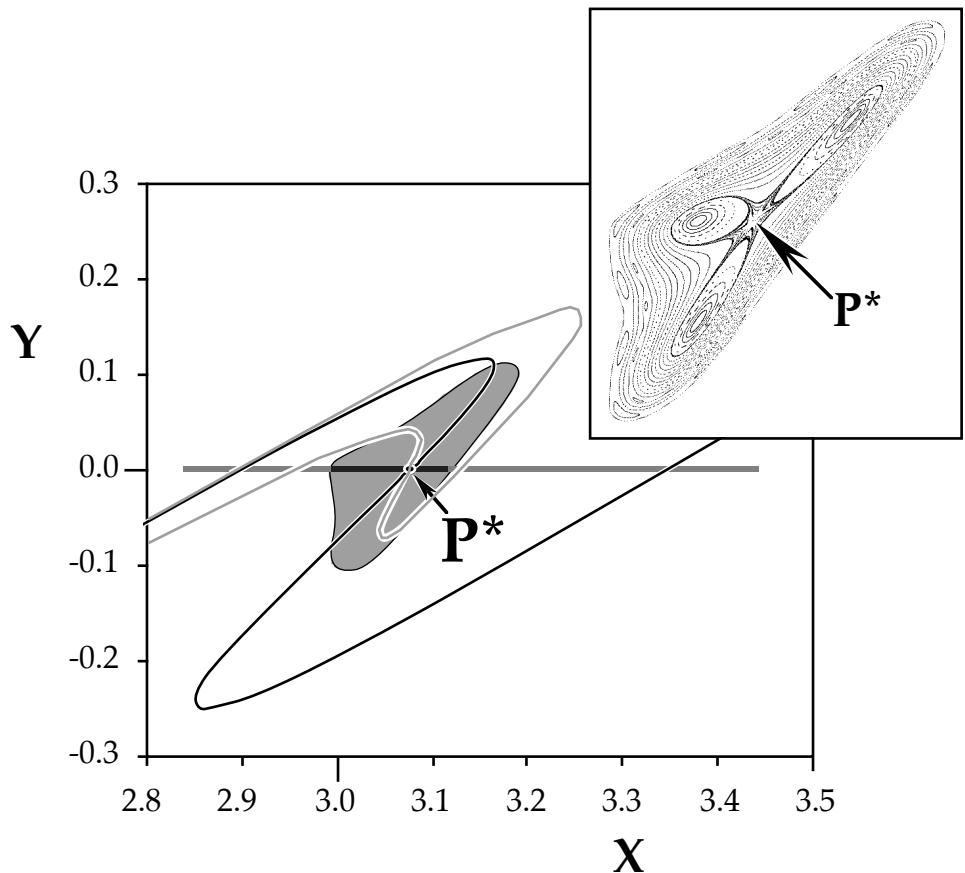

**Figure 3 - Folds created near unstable fixed point, $P^*$, of standard map. Inset: resulting coherent structure.**

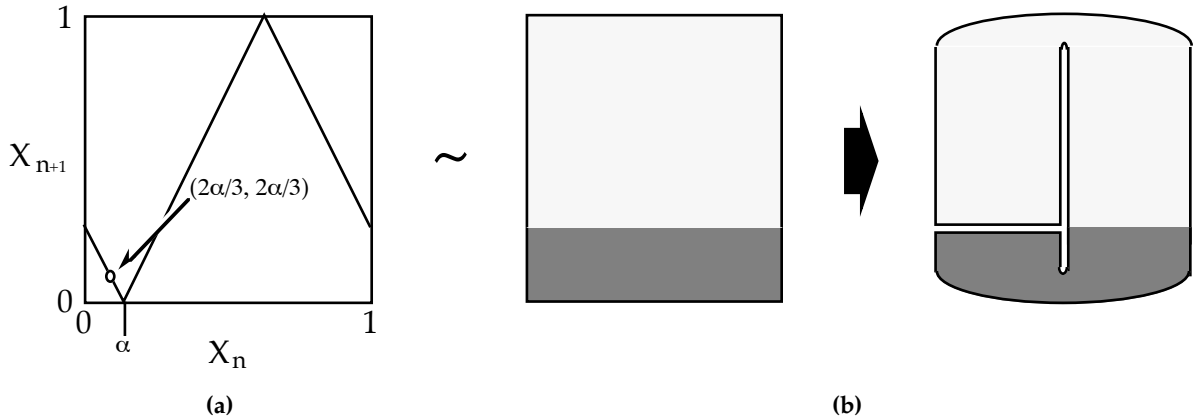

**Figure 2 - Double fold tent map (a) and a corresponding double horseshoe map (b)**

The analogous operation for a 2-D horseshoe is to create a second fold which folds the bottom portion of the original sheet back onto itself. This is shown in Fig. 2(b), where a rectangle is stretched and then both a bottom portion and an upper portion are folded over to produce the double horseshoe map shown in the right of Fig. 2. Points initially in either the lower (dark grey) or the upper (light grey) portion then stay in that portion after successive applications of this map. This observation constitutes the crucial contribution of this letter, for horseshoe maps are widely used as a model for the stretching and folding produced in chaotic systems[8].

The preceding examples portray, in an uncomplicated setting, a recipe for creating coherent structures by introducing folds near fixed points. In more realistic problems, complications can be expected; nevertheless one can create coherent structures in such systems by following the same recipe. For example, consider the standard map, defined by:

$$Y_{n+1} = Y_n - K \cdot \sin(X_n)$$
$$X_{n+1} = X_n + Y_{n+1} , \qquad [3]$$

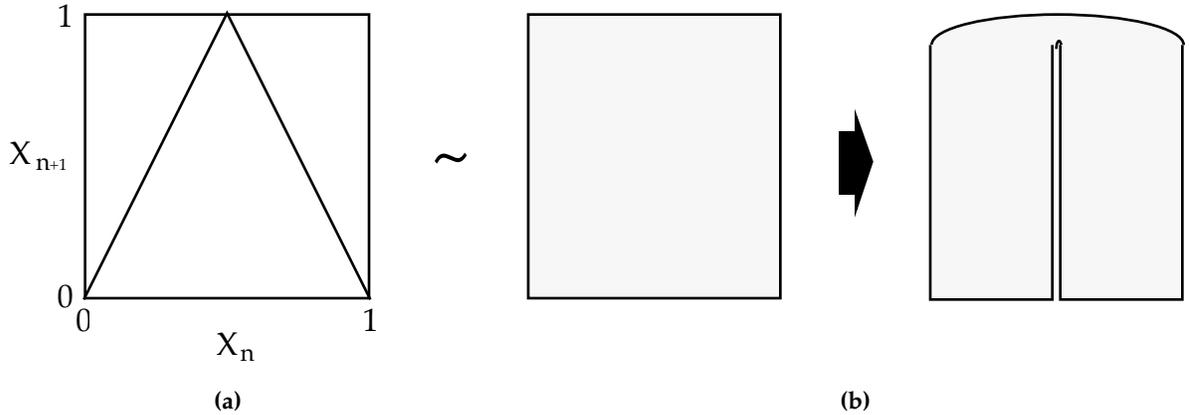

**Figure 1 - Single fold tent map (a) and corresponding horseshoe map (b)**

Almost every small interval between 0 and 1 is stretched by a factor of two after every iteration of the map. It is therefore *prima facie* impossible for any such interval to contain a localized structure. The only exception to this rule occurs near a fold, where an interval is doubled over, thus conceivably counteracting the stretching. To construct an invariant interval, *I*, by using folds, we must place a fold sufficiently near a fixed point that points within *I* remain in *I* after successive iterations of the map.

In the case of the tent map, it is easy to achieve this by changing the phase, $\alpha$, as defined in the new map:

$$X_{n+1} = \begin{cases} 2 \cdot (\alpha - X_n) & \text{if } 0 \leq X_n < \alpha \\ 2 \cdot (X_n - \alpha) & \text{if } \alpha \leq X_n < (1/2 + \alpha) \\ 2 \cdot (1 + \alpha - X_n) & \text{if } (1/2 + \alpha) \leq X_n \leq 1 \end{cases} \quad [2]$$

In this map, shown in Fig. 2(a), the interval $I = (0, 2\alpha)$ now contains both a fixed point, at $X_{n+1} = X_n = 2\alpha/3$, and a fold, at $X_n = \alpha$, and points initially in *I* remain forever in *I* after repeated iterations of the map [2]. We emphasize that both maps [1] and [2] are chaotic throughout the unit interval. The effect of introducing the fold is thus strictly limited to constraining trajectories to the selected region.

Coherent structures are remarkably ubiquitous in a variety of natural problems. Examples range from planetary fluid dynamics[1] (e.g. Jupiter's red spot) and chaotic advection[2] to turbulent boundary layer dynamics[3] and localized behavior in coupled oscillator arrays[4]. One of the most surprising aspects of these regular, localized structures is that they seem to be especially prevalent in irregular, turbulent dynamical systems. Moreover, research over the past four decades has demonstrated that many extended systems can be well represented by a limited number of properly selected basis functions[5]. This implies that low dimensional and localized dynamics can exist in formally infinite dimensional extended systems -- such as turbulent fluids.

In this letter, we study the nature of coherent structures in chaotic systems in terms of simple conceptual tools. We address two questions, both in the context of area preserving flows. First, we ask why coherent structures are prevalent. Second, we ask why coherent structures in chaotic flows are not torn apart by the stretching and folding intrinsic to chaos[6]. To approach these issues, we first develop a method for creating a coherent structure in a uniformly chaotic 1-D map. We then show that this method can be extended to more complicated, 2-D maps. Finally we apply the method to the chaotic motion of a periodically forced fluid.

The key idea is to intentionally place folds of horseshoes near periodic points. The simplest version of the idea can be illustrated in the tent map (Fig. 1(a)):

$$X_{n+1} = \begin{cases} 2 \cdot X_n & \text{if } 0 \leq X_n < 1/2 \\ 2 \cdot (1-X_n) & \text{if } 1/2 \leq X_n \leq 1 \end{cases} \quad [1]$$

Such a map is the 1-D analog of a 2-D horseshoe map[7]. In Fig. 1(b) we show a horseshoe map produced by stretching a rectangle, shown in the center of Fig. 1, and then folding the stretched rectangle in half to produce the horseshoe shape shown in the right of Fig. 1.

# Using Horseshoes to Create Coherent Structures

**Troy Shinbrot** and **J.M. Ottino**
*Laboratory for Fluid Mechanics, Chaos and Mixing*
*Dept. of Chemical Engineering*
*Northwestern University*
*Evanston, IL  60208*

In this letter, we show that coherent structures are related to folds of horseshoes which are present in chaotic systems.  We develop techniques that allow us to  construct coherent structures by manipulating folds in three prototypical problems: a 1-D chaotic map, a 2-D chaotic map, and a chaotically advected fluid.  The ability to  construct such structures is of practical importance for the control of chaotic or turbulent extended systems such as fluids, plasmas, and coupled oscillator arrays.